\documentclass{article} 
\newcommand{\ie}{{\it i.e.}}
\newcommand{\pc}{p_{\rm c}}
\newcommand{\bea}{\begin{eqnarray}}
\newcommand{\eea}{\end{eqnarray}}
\newcommand{\f}{\frac}
\newcommand{\no}{\nonumber}

\usepackage{graphicx}
\usepackage{amssymb}

\font\ffour=cmss10

\begin{document}
\vspace{2cm}

\begin{center}
{\large \bf The critical point of $k$-clique percolation in the 
Erd\H os-R\'enyi graph}
\vspace{2cm}

Gergely Palla$^{\dagger\ddagger}$, Imre Der\'enyi$^{\ddagger}$ and 
Tam\'as Vicsek$^{\dagger\ddagger}$
\vspace{1cm}

$^{\dagger}$Biological Physics Research Group of HAS, P\'azm\'any P.\ stny.\ 1A, H-1117 Budapest, Hungary, \\
$^{\ddagger}$ Dept. of Biological Physics, E\"otv\"os University,
 P\'azm\'any P.\ stny.\ 1A, H-1117 Budapest, Hungary.
\end{center}

\newpage

\begin{abstract}
Motivated by the success of a $k$-clique percolation method
 for the identification of overlapping communities in large real networks,
 here we study the $k$-clique percolation problem in the Erd\H{o}s-R\'enyi 
graph.
When the probability $p$ of two nodes being connected is above a 
 certain threshold $p_c(k)$, the complete subgraphs of size $k$
 (the $k$-cliques)
 are organized into a giant cluster.
By making some assumptions that are expected to be valid below the
threshold, we determine the average size of the $k$-clique percolation
clusters, using a generating function formalism. From the divergence of
this average size we then derive an analytic expression for the
critical linking probability $p_c(k)$.
\end{abstract}

\section{Introduction}

Many complex systems in nature and society can be successfully represented
in terms
of networks capturing the intricate web of connections among the units
they are made of. Graphs corresponding
 to these real networks exhibit unexpected non-trivial properties,
{\it e.g.}, new kinds of degree distributions, anomalous diameter,
spreading phenomena, clustering coefficient, and correlations
\cite{watts-strogatz,barabasi-albert,albert-revmod,dm-book,barrat}.
In recent years, there has been a quickly growing interest in
the local structural
units of networks. Small and well defined subgraphs consisting
of a few vertices have been introduced as ``motifs'' \cite{alon}.
Their distribution and clustering properties
\cite{alon,vazquez-condmat,kertesz-condmat}
can be considered as important global
characteristics of real networks.
Somewhat larger units, associated with more highly interconnected parts
\cite{domany-prl,gn-pnas,zhou,newman-fast,al-parisi,huberman,spektral,potts,scott-book,pnas-suppl,everitt-book,knudsen-book,newman-europhys}
 are usually called
clusters, communities, cohesive groups, or modules, with no widely accepted,
 unique definition. Such building blocks (functionally related  proteins \cite{ravasz-science,spirin-pnas}, industrial sectors \cite{onnela-taxonomy},
groups of people \cite{scott-book,watts-dodds}, cooperative players
\cite{play1,play2}, {\it etc.}) can play a crucial role in the structural and functional
properties of the networks involved. The presence of
 communities is also a relevant and
informative signature of the hierarchical nature of complex systems
\cite{ravasz-science,vicsek-nature,self-sim-coms}.

Most of the methods used for the
identification of communities rely on dividing the network into smaller
pieces. The biggest drawback of these methods is that they do not allow
{\it overlapping} for the communities. On the other hand, the communities 
in a complex system are often not isolated from each other, 
 but rather, they have overlaps,
 {\it e.g.}, a protein can be part of more than one functional
unit \cite{protein-complexes}, and people can be members in different social
groups at the same time \cite{wasserman}. One possibility to overcome
 this problem is to use a community definition based on $k$-clique 
percolation \cite{our-PRL,our-Nature}. In this approach the communities
 are associated with $k$-clique percolation clusters, and can overlap
 with each other. The communities of large real networks obtained
 with this method were shown to have significant overlaps, and the
statistical properties of the  communities exhibited non-trivial universal
 features \cite{our-Nature}. 

In this manuscript we focus on the basic properties
of $k$-clique percolation.
In a recent work we have already proposed an expression for the critical point
 of the $k$-clique percolation  in the Erd\H os-R\'enyi (E-R) graph
 using simple heuristic arguments \cite{our-PRL}. This expression has also
 been supported by our numerical simulations.
The goal of this manuscript is to make these result stronger by
providing a more detailed analytical derivation using only a few
reasonable assumptions, expected to be valid below the critical point.
We note that the 
 critical point of $k$-clique percolation plays a crucial role
 in the community finding as well. When dealing
 with a network containing weighted links, one can introduce 
a weight threshold and exclude
 links weaker than the threshold from the investigation. When the threshold
 is very high, only a few disintegrated community remains, whereas in  case
 of a very low threshold, a giant community arises smearing out the 
details of the community structure by merging (and making invisible) 
many smaller communities. To find a community 
structure as highly structured as possible, one needs to set the threshold
 close to the critical point of the $k$-clique percolation.
 
\section{$k$-clique percolation in the E-R graph}

In the field of complex networks, the classical E-R uncorrelated random graph
\cite{e-r} serves both as a test bed for checking all sorts of new ideas 
concerning networks in general, and
as a prototype of random graphs to which all other random graphs can be
compared. One of the most conspicuous early
result on the E-R graphs was related to the percolation transition
taking place at $p=\pc \equiv 1/N$, where $p$ is the probability that
two vertices  are connected by an edge and $N$ is the total number
of vertices in the graph. 
The appearance of a {\em giant component} in a
network, which is also referred to as the {\em percolating component},
results in a dramatic change in the overall topological features
of the graph and has been in the center of interest for
other networks as well.

In this manuscript we address the general question of subgraph
percolation in the E-R model. In particular, we provide 
an analytic expression for the
threshold probability at which the percolation transition of
complete subgraphs of size $k$ (the $k$-cliques) takes place.
\begin{figure}[b!]
\centerline{\includegraphics[angle=0,width=0.85\columnwidth]{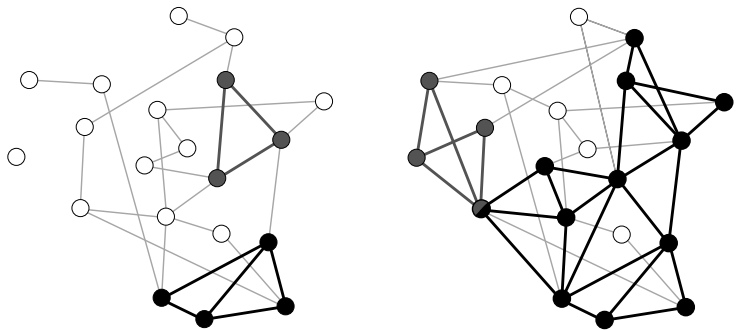}}
\caption{\ffour Sketches of two E-R graphs of $N=20$ vertices and with edge
probabilities $p=0.13$ (left one) and $p=0.22$ (right one, generated by
adding more random edges to the left one). In both cases all the edges
belong to a ``giant'' connected component, because the edge
probabilities are much larger than the threshold ($\pc \equiv 1/N =
0.05$) for the classical E-R percolation transition. However, in the
left one $p$ is below the 3-cliques percolation threshold, $\pc(3)
\approx 0.16$, calculated from Eq.\ (\ref{pcrit}), therefore, only a
few scattered 3-cliques (triangles) and small 3-clique percolation
clusters (distinguished by black and dark gray edges) can be observed.
In the right one, on the other hand, $p$ is above this threshold and,
as a consequence, most 3-cliques accumulate in a ``giant'' 3-clique
percolation cluster (black edges). This graph also illustrates the
overlap (half black, half dark gray vertex) between two clusters
(black and dark gray).
}
\label{fig_sketch}
\end{figure}
Before proceeding we need to go through  some basic definitions:
\begin{itemize}
 \item {\em $k$-clique}: a complete (fully connected)
subgraph of $k$ vertices \cite{bollobas-book}.
\item {\em $k$-clique adjacency}: two $k$-cliques are adjacent if they
share $k-1$ vertices, \ie, if they differ only in a single node.
\item {\em $k$-clique chain}: a subgraph, which
is the union
of a sequence of adjacent $k$-cliques.
\item {\em $k$-clique connectedness}: two $k$-cliques are
$k$-clique-connected if they are parts of a $k$-clique chain.
\item {\em $k$-clique percolation cluster (or component)}: a
maximal $k$-clique-connected subgraph, \ie, it
is the union
of all $k$-cliques that are $k$-clique-connected to a particular
$k$-clique. 
\end{itemize}
The above concept of $k$-clique percolation is illustrated 
in Fig.\ref{fig_sketch}, where both
graphs contain two 3-clique percolation clusters, the smaller ones in
dark gray and the larger ones in black.
We note that these objects can be considered as interesting
specific cases of the general graph theoretic objects defined
by Everett and Borgatti
\cite{clique-overlap}
and by Batagelj and Zaversnik
\cite{short-cycles}
in very different contexts.

An illustration of the $k$-clique percolation clusters can be given 
by ``{\it $k$-clique template rolling}''. A $k$-clique template can be 
thought of as an object that
 is isomorphic to a complete graph of $k$ nodes. Such a template can be
 placed onto any $k$-clique of the network, and rolled to an
adjacent $k$-clique by relocating one
of its nodes and keeping its other $k-1$ nodes fixed. Thus, the
$k$-clique-communities of a graph are all those subgraphs that
can be fully explored by rolling a $k$-clique template in them but cannot
be left by this template. We note that a $k$-clique percolation 
cluster is very much like a regular edge
percolation cluster in the
{\em $k$-clique adjacency graph}, where the vertices represent the
$k$-cliques of the original graph, and there is an edge between two
vertices if the corresponding $k$-cliques are adjacent. Moving a
particle from one vertex of this adjacency graph to another one along
an edge is equivalent to rolling a $k$-clique template from one
$k$-clique of the original graph to an adjacent one.

\section{The generating functions}

In our investigation of the critical point of the $k$-clique 
percolation in the E-R graph we shall rely on the generating
 function formalism in a fashion similar to that of Ref.\cite{newman-arbitPk}.
 Therefore, in this section we first summarize the definition and the 
most important properties of the generating functions. If a random 
variable $\xi$  
can take non-negative integer values according to some
probability distribution ${\cal P}(\xi=n)\equiv\rho(n)$,
 then the corresponding generating function
is given by
\bea
G_{\rho}(x)\equiv\left<x^{\xi}\right>=\sum_{n=0}^{\infty}\rho(n)x^n.
\eea
The generating-function of a properly normalized distribution 
is absolute convergent for all $|x|\leq1$ and hence has no singularities
in this region. For $x=1$ it is simply
\bea
G_{\rho}(1)=\sum_{n=0}^{\infty}\rho(n)=1.
\eea
The original probability distribution and its moments can be 
obtained from the generating-function as
\bea
\rho(n)&=&\f{1}{n!}\left.\f{d^nG_{\rho}(x)}{dx^n}\right|_{x=0},
\label{eq:vissza}\\
\left<\xi^l\right>&=&\sum_{n=0}^{\infty}n^l\rho(n)=\left[\left(x\f{d}{dx}
\right)^lG_{\rho}(x)\right]_{x=1}.\label{eq:moment}
\eea
And finally, if $\eta=\xi_1+\xi_2+...+\xi_l$, where $\xi_1,\xi_2,...,\xi_l$ are
independent random variables (with non-negative integer values), 
then the generating function corresponding to 
${\cal P}(\eta=n)\equiv\sigma(n)$ is given by
\bea
G_{\sigma}(x)&=&\left<x^{\eta}\right>=\left<x^{\xi_1}x^{\xi_2}\cdots x^{\xi_l}\right>=
\left<x^{\xi_1}\right>\left<x^{\xi_2}\right>\cdots\left<x^{\xi_l}\right>=
\no \\ & &
G_{\rho_1}(x)G_{\rho_2}(x)\cdots G_{\rho_l}(x). \label{eq:powers}
\eea

\section{The critical point}

In this section we arrive at the derivation of the critical point of the 
$k$-clique percolation in the E-R graph in the $N\rightarrow\infty$ limit. 
We begin by considering
 the probability distribution $r(n)$ of the number of $k$-cliques adjacent
 to a randomly selected $k$-clique. Finding a $k$-clique $B$ adjacent to 
 a selected $k$-clique $A$ is equivalent to finding a node 
 outside $A$ linked to at least $k-1$ nodes in $A$. 
 The number of possibilities for this node is $N-k$. 
Links in the E-R graph are independent
 of each other, therefore the probability that a given node is linked
 to all nodes in $A$ is $p^k$, whereas the probability that it is linked
 to $k-1$ nodes in $A$ is $k(1-p)p^{k-1}$. Therefore, to leading order in $N$
 the average number of $k$-cliques adjacent to a randomly selected one is
 \bea
 \left<r\right>=(N-k)\left[k(1-p)p^{k-1}+p^k\right]\simeq Nkp^{k-1}.
 \eea
 From the independence of the links  it also follows that 
 the probability distribution $r(n)$ becomes
 Poissonean, which can be written as
\bea
r(n)=\exp\left(-Nkp^{k-1}\right)\f{\left(Nkp^{k-1}\right)^n}{n!}.
\eea

\begin{figure}[hbt]
{\centerline{\includegraphics[width=0.85\textwidth]{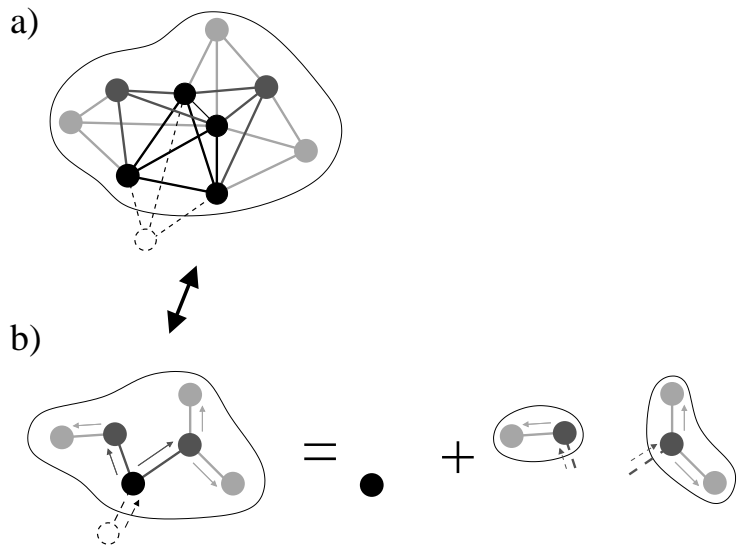}}}
\caption{\ffour Schematic picture of the evaluation of the size of 
a $k$-clique percolation cluster by counting the number of $k$-cliques
 reached in an ``invasion'' process. a) Let us suppose that we arrived
 at the black colored $k$-clique from
 the $k$-clique marked by dashed lines. In the next step we must 
 proceed to the $k$-cliques shown in dark gray, and then finally to the
 $k$-cliques marked in light gray. b) The corresponding $k$-clique adjacency
 graph is shown on the right. The size of the connected component
 in the $k$-clique adjacency graph we can invade from the black $k$-clique
 (by excluding the link through which we initially reached it) 
is equal to one plus
 the sum of the sizes of the connected components invaded from the dark
 gray $k$-cliques in the same way.
\label{fig:gen_func_meth_magy}} 
\end{figure}

Let us suppose that we are below the percolation threshold, therefore,
 $k$-cliques are rare, adjacent $k$-cliques are even more rare, and
 loops in the $k$-clique adjacency graph are so rare that we can assume
 it to be tree-like\cite{assumption}. In this case the size of a 
connected component in the $k$-clique adjacency graph (corresponding
 to a $k$-clique percolation cluster) can be evaluated by  
 counting the number of $k$-cliques reached in an ``invasion'' process
 as follows. We start at an arbitrary $k$-clique in the component, and in 
 the first step we invade all its neighbors in the $k$-clique adjacency graph.
 From then on, whenever a $k$-clique is 
 reached, we proceed by invading all its neighbors, except for the one 
the $k$-clique has been reached from, 
as shown in Fig.\ref{fig:gen_func_meth_magy}a.
 In terms of the original graph, this 
 is equivalent to rolling a $k$-clique template to all adjacent
 $k$-cliques except for the one we arrived from in the previous step. 

In the invasion process described above, we can assign to 
each $k$-clique the subgraph in the $k$-clique percolation cluster 
that was invaded from it. 
(Note that we assumed the $k$-clique adjacency graph to be tree-like).
Let us denote
 by $I(n)$ the probability, that the subgraph reached from an arbitrary 
starting $k$-clique in the invasion process contains $n$ 
number of $k$-cliques, 
including the starting $k$-clique as well. This subgraph
 is actually equal to a $k$-clique percolation cluster.
Similarly, let $H(n)$ denote the probability that the subgraph invaded
 from a $k$-clique appearing later in the invasion process 
({\it i.e.}, from a $k$-clique that is not the starting one) contains
 $n$ number of $k$-cliques. 
This is equivalent to the probability 
 that by starting at a randomly selected $k$-clique
 and trying to roll a $k$-clique template via all possible subsets 
 of size $k-1$ except for one, then by succedingly rolling the template 
on and on, in all possible directions without turning back,
 a $k$-clique percolation ``branch'' of size $n$ is invaded. 
 And finally, let $H_{m}(n)$ be the probability, 
 that if pick $m$ number of $k$-cliques randomly, then the sum of the 
sizes of the $k$-clique branches that we can invade in this way consists of
 $n$ number of $k$-cliques. Since we are below the percolation threshold,
 the $k$-clique adjacency graph consists of many dispersed components of
 small size, and the probability that two (or more) $k$-cliques out of 
 $m$ belong to the same $k$-clique percolation cluster is negligible.
 Hence, according to Eq.(\ref{eq:powers}), 
the generating functions corresponding to $H(n)$ and $H_m(n)$,
 denoted by $G_H(x)$ and $G_{H_m}(x)$ respectively are related to each other
 as:
\bea
G_{H_m}(x)=\left[G_{H}(x)\right]^m.\label{eq:G_rel}
\eea

Let $q(n)$ denote the probability, that for a randomly selected
$k$-clique, by excluding one of its possible subsets of size $k-1$,
 we can roll a $k$-clique template through the remaining subsets to $n$ 
adjacent $k$-cliques. This distribution is very similar
 to $r(n)$, except that in this case we can use only $k-1$ subsets instead of
 $k$ in the $k$-clique to roll the $k$-clique template further, therefore
\bea
q(n)=\exp\left(-N(k-1)p^{k-1}\right)\f{\left(N(k-1)p^{k-1}\right)^n}{n!}.
\eea
 By neglecting the loops in 
the $k$-clique adjacency graph, $H_n$ can be expressed as
\bea
H(n)=q(0)H_{0}(n-1)+q(1)H_{1}(n-1)+q(2)H_{2}(n-1)+... ,
\label{eq:H_k_main}
\eea
as explained in Fig.\ref{fig:gen_func_meth_magy}b.
 By taking the generating function of both sides and using 
Eqs.(\ref{eq:vissza}) and (\ref{eq:G_rel}), we obtain
\bea
G_H (x)&=&\sum_{n=0}^{\infty}\left[\sum_{m=0}^{\infty}
q(m)H_{m}(n-1)\right]x^n=\nonumber \\ &=&\sum_{n=0}^{\infty}\left[\sum_{m=0}^{\infty}
q(m)\f{1}{(n-1)!}\left.\f{d^{n-1}}{dx^{n-1}}\left[
G_H(x)\right]^m\right|_{x=0}\right]x^n=\nonumber \\
&=&\sum_{m=0}^{\infty}q(m)\left[G_H(x)\right]^mx
=xG_q\Biggl(G_H(x)\Biggr), \label{eq:G_H}
\eea
where $G_q(x)$ denotes the generating function of the distribution 
$q(n)$.

We can write
an equation similar to Eq.(\ref{eq:H_k_main}) for $I(n)$ as 
 well, in the form of
\bea
I(n)=r(0)H_{0}(n-1)+r(1)H_{1}(n-1)+r(2)H_{2}(n-1)+\dots
\eea
Again, by taking the generating functions of both sides we arrive at
\bea
G_I(x)=
xG_r(G_H(x)), \label{eq:G_I}
\eea
where $G_r(x)$ denotes the generating function of $r(n)$.
From Eqs.(\ref{eq:moment}) and (\ref{eq:G_I}) we get 
\bea
\left<I \right>=G_I'(1)=G_{r}(G_H(1))+
G'_{r}(G_H(1))G_H'(1)=1+G'_{r}(1)G_H'(1) \label{eq:I_n_av}
\eea
for the average size of the components invaded from a randomly selected
 $k$-clique.
Using Eq.(\ref{eq:G_H}) we can write
\bea
G_H'(1)&=&G_{q}(G_H(1))+G_{q}'(G_H(1))G_H'(1)=1+G_{q}'(1)
G_H'(1),
\eea
from which $G_H'(1)$ can be expressed as
\bea
G_H'(1)=\f{1}{1-G_{q}'(1)}.
\eea
By substituting this back into Eq.(\ref{eq:I_n_av}) we get
\bea
\left<I \right>=1+\f{G'_{r}(1)}{1-G_{q}'(1)}=1+
\f{\left<r\right>}{1-\left<q\right>}.
\eea
The above expression for the expected size of the connected components 
in the $k$-clique adjacency graph invaded from a randomly selected
 $k$-clique diverges when
\bea
\left<q\right>=N(k-1)p^{k-1}=1.
\label{eq:G_q'=1}
\eea
This point marks the phase transition at which a giant component
(corresponding to a giant $k$-clique percolation cluster) first appears. 
Therefore, our final result for the critical linking probability  for
 the appearance of the giant component is
\begin{equation}
\pc(k)={1\over {[N(k-1)]^{1\over {k-1}}}}.
\label{pcrit}
\end{equation}
This result reassures the findings of \cite{our-PRL} based on heuristic
 arguments and numerical simulations. Obviously, for $k=2$ our result agrees 
with the known percolation threshold ($\pc=1/N$) for E-R graphs, 
because 2-clique connectedness is equivalent to regular (edge) connectedness. 

\section{Conclusions}
The phenomenon of $k$-clique percolation provides an effective tool 
for finding overlapping communities in large networks. 
In this article we derived the critical linking 
 probability for the E-R graph in the $N\rightarrow\infty$ limit. Our
 method involved the use of generating functions and was based on the 
 assumption that up to the critical point, loops
 in the $k$-clique adjacency graph are negligible. Our findings are in
 complete agreement with earlier results based on heuristic arguments
 and numerical simulations.  

This work has been supported in part by
the Hungarian Science Foundation (OTKA), grant Nos. F047203 and T049674.


\begin{thebibliography}{}

\bibitem{watts-strogatz}
D. J. Watts and S. H. Strogatz,   
\textit{Nature} \textbf{393}, 440 (1998).

\bibitem{barabasi-albert}
A.-L. Barab{\'a}si and R. Albert,
\textit{Science} \textbf{286}, 509 (1999).

\bibitem{albert-revmod}
R. Albert and A.-L. Barab{\'a}si,
\textit{Rev.\ Mod.\ Phys.} \textbf{74}, 47 (2002).

\bibitem{dm-book}
J. F. F. Mendes and S. N. Dorogovtsev,
\textit{Evolution of Networks: From Biological Nets to the Internet and WWW}
(Oxford University Press, Oxford, 2003).

\bibitem{barrat}
A. Barrat, M. Barthelemy, and A. Vespignani,
\textit{Phys.\ Rev.\ Lett.} \textbf{92}, 228701 (2004).

\bibitem{alon}
R. Milo, S. Shen-Orr, S. Itzkovitz, N. Kashtan, D. Chklovskii, and U. Alon,
\textit{Science} \textbf{298}, 824 (2002).

\bibitem{vazquez-condmat}
A. Vazquez, R. Dobrin, D. Sergi, J.-P. Eckmann, Z. Oltvai and
A.-L. Barab{\'a}si,
\textit{Proc.\ Natl.\ Acad.\ Sci.\ USA} \textbf{101}, 17945 (2004).

\bibitem{kertesz-condmat}
J.-P. Onnela, J. Saram{\"a}ki, J. Kert{\'e}sz, and K. Kaski,
\textit{Phys.\ Rev.\ E} \textbf{71}, 065103 (2005).



\bibitem{domany-prl}
M. Blatt, S. Wiseman, and E. Domany,
\textit{Phys.\ Rev.\ Lett.} \textbf{76}, 3251 (1996).

\bibitem{gn-pnas}
M. Girvan and M. E. J. Newman,
\textit{Proc.\ Natl.\ Acad.\ Sci.\ USA} \textbf{99}, 7821 (2002).

\bibitem{zhou}
H. Zhou,
\textit{Phys.\ Rev.\ E} \textbf{67}, 061901 (2003).

\bibitem{newman-fast} 
M. E. J. Newman,
\textit{Phys.\ Rev.\ E} \textbf{69}, 066133 (2004).

\bibitem{al-parisi}
F. Radicchi, C. Castellano, F. Cecconi, V. Loreto, and D. Parisi,
\textit{Proc.\ Natl.\ Acad.\ Sci.\ USA} \textbf{101}, 2658 (2004).

\bibitem{spektral}
L. Donetti and M. A. Mu{\~n}oz,
\textit{J.\ Stat.\ Mech.} (2004) P10012.

\bibitem{huberman}
D. M. Wilkinson and B. A. Huberman,
\textit{Proc.\ Natl.\ Acad.\ Sci.\ USA} \textbf{101}, 5241 (2004).

\bibitem{potts}
J. Reichardt and S. Bornholdt,
\textit{Phys.\ Rev.\ Lett.} \textbf{93}, 218701 (2004).


\bibitem{scott-book}
J. Scott,
\textit{Social Network Analysis: A Handbook}, 2nd ed.
(Sage Publications, London, 2000).

\bibitem{pnas-suppl}
Shiffrin, R. M. \& B{\"o}rner, K.
Mapping knowledge domains.
\textit{Proc.\ Natl.\ Acad.\ Sci.\ USA} \textbf{101}, 5183 Suppl.\ 1 (2004).

\bibitem{everitt-book}
Everitt, B. S.
%
\textit{Cluster Analysis}, 3th ed.
(Edward Arnold, London, 1993).

\bibitem{knudsen-book}
Knudsen, S.
%
\textit{A Guide to Analysis of DNA Microarray Data}, 2nd ed.
(Wiley-Liss, 2004).

\bibitem{newman-europhys}
Newman, M. E. J.
Detecting community structure in networks.
\textit{Eur. Phys. J. B}, \textbf{38}, 321 (2004).


\bibitem{ravasz-science}
E. Ravasz,  A. L. Somera, D. A. Mongru, Z. Oltvai and A.-L. 
Barab\'asi
\textit{Science} \textbf{297}, 1551 (2002).

\bibitem{spirin-pnas}
V. Spirin and L. A. Mirny
\textit{Proc.\ Natl.\ Acad.\ Sci.\ USA} \textbf{100} 12123 (2003).

\bibitem{onnela-taxonomy}
J. P. Onnela, A. Chakraborti, K. Kaski, J. Kert{\'e}sz and A. Kanto
\textit{Phys.\ Rev.\ E} \textbf{68}, 056110 (2003).

\bibitem{watts-dodds}
D. J. Watts, P. S. Dodds and M. E. J. Newman
\textit{Science} \textbf{296}, 1302 (2002).

\bibitem{play1}
J. Vukov and Gy. Szab\'o
\textit{Phys.\ Rev.\ E} \textbf{71}, 036133 (2005).

\bibitem{play2}
Gy. Szab\'o, J. Vukov and A. Szolnoki
\textit{Phys.\ Rev.\ E} \textbf{72} 047107 (2005).

\bibitem{vicsek-nature}
T. Vicsek
\textit{Nature} \textbf{418}, 131 (2002).

\bibitem{self-sim-coms}
R. Guimer\'a, L. Danon, A. D{\'\i}az-Guilera, F. Giralt and A. Arenas
\textit{Phys. Rev. E} \textbf{68}, 065103 (2003).

\bibitem{protein-complexes}
A. C. Gavin \textit{et al.}
\textit{Nature} \textbf{415}, 141 (2002).

\bibitem{wasserman}
\textit{P. Eds Carrington and J.  Scott and S. Wasserman}
\textit{Models and Methods in Social Network Analysis}, Ch. 7,
Edited by K. Faust, 
Cambridge University Press, New York (2005).

\bibitem{our-PRL}
I. Der\'enyi, G. Palla and T. Vicsek
\textit{Phys. Rev. Lett.} \textbf{94}, 160202 (2005).


\bibitem{our-Nature}
G. Palla, I. Der\'enyi, I. Farkas and T. Vicsek
\textit{Nature} \textbf{435}, 814 (2005).



\bibitem{e-r}
P. Erd{\H{o}}s and A. R{\'e}nyi,
\textit{Publ.\ Math.\ Inst.\ Hung.\ Acad.\ Sci.} \textbf{5}, 17 (1960).

\bibitem{bollobas-book}
B. Bollob{\'a}s,
\textit{Random graphs}, 2nd ed.
(Cambridge University Press, Cambridge, 2001).

\bibitem{clique-overlap}
M. G. Everett and S. P. Borgatti,
\textit{Connections} \textbf{21}, 49 (1998).

\bibitem{short-cycles}
V. Batagelj and M. Zaversnik,
arXiv:cs.DS/0308011 (2003).


\bibitem{newman-arbitPk}
M. E. J. Newman, S. H. Strogatz, and D. J. Watts,
\textit{Phys.\ Rev.\ E} \textbf{64}, 026118 (2001).

\bibitem{assumption}
This assumption is an approximation since the adjacency graph is
weakly assortative.






\end{thebibliography}
\end{document}